\documentclass[seceq,supplement]{ptptex}

\usepackage{graphicx}
\usepackage{wrapft}

\title{
Non-equilibrium Relaxation Analysis on Two-dimensional Melting%
}

\author{
Hiroshi \textsc{Watanabe}%
}

\inst{
Department of Complex
Systems Science, Graduate School of Information Science,
Nagoya University, Furouchou, Chikusa-ku, Nagoya 464-8601, Japan
}

\abst{
The melting transition in the hard-disk system is considered.
Non-equilibrium relaxation analysis of the six-fold bond-orientational 
order parameter has been carried out.
The critical point between the hexatic and the fluid phase is determined 
on the basis of the dynamic scaling hypothesis.
The value of the critical exponent $\eta$ is 
determined from the fluctuation of the order parameter at the criticality
as $\eta = 0.25(2)$ which is consistent with the prediction by the 
Kosterlitz-Thouless-Halperin-Nelson-Young theory.
}

\renewcommand{\v}[1]{{\bf #1}}
\newcommand{\ave}[1]{\left< #1 \right>}

\begin{document}

\maketitle

\section{Introduction}

\begin{figure}[bt]
\begin{center}
\includegraphics[width=0.5\linewidth]{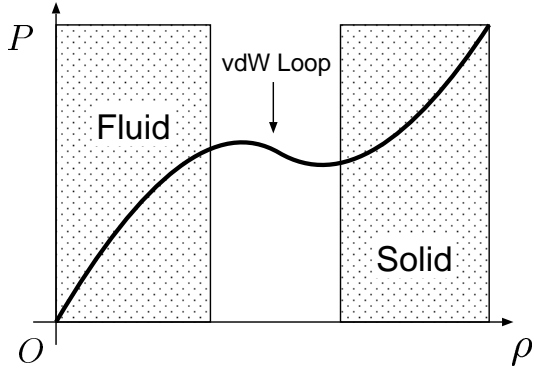}
\end{center}
\caption{
The density-pressure diagram of the particle-system only with the repulsive interaction.
The melting behavior can be confirmed because of the 
existence of the van der Waals loop.
Note that, this figure is exaggerated. The domain of the loop
becomes very small as the system size becomes larger.
}
\label{fig_pressure}
\end{figure}

In 1957, Alder {\it et al.} reported the melting transition
in the hard-sphere system\cite{Alder1957}.
The fact that particles only with repulsive interaction
involve the phase transition has surprised many researchers.
This transition was also confirmed in the hard-disk system~\cite{Alder1962},
and these kinds of transitions are now often called the Alder transition.
The Alder transition seems to be the usual first-order type
since it is characterized by a van der Waals loop like 
behavior in the pressure-density diagram (see Fig.~\ref{fig_pressure}).
However, Mermin has proved that the long-range correlation of the translational order
is inhibited\cite{Mermin}, and therefore, the two-dimensional melting
cannot be the usual order-disorder phase transition.
Halperin, Nelson and Young
proposed the theory of the two-dimensional melting based on the 
unbinding mechanism of the defects~\cite{HalperinNelson, Young}.
They have suggested that the two-dimensional melting involves
the Kosterlitz-Thouless transition twice.
This theory is now often referred to the Kosterlitz-Thouless-Halperin-Nelson-Young (KTHNY) theory.
Competing theory predicting the first-order transition was proposed by Chui~\cite{Chui}.
Many experimental and numerical studies have followed in order to clarify the
nature of the transition.
In spite of these efforts, the strong finite-size effect has kept us from fixing this problem
and the nature of the two-dimensional melting remains unanswered~\cite{KJ}.

In this paper, an overview of the two-dimensional melting is given first.
After the Mermin's theorem and the KTHNY theory are described,
a new method, which utilizes  the non-equilibrium relaxation behavior of the order parameter,
is introduced in order to avoid the finite-size effect.
Conclusions and some recent developments are summarized in the final section.

\section{The KTHNY theory}

\subsection{Translational order}

Roughly speaking, states of matter can be classified into two phases, fluid and solid.
Fluid is a substance which deforms under an infinitesimal shear stress,
while solid requires finite stress.
In this definition, gas and liquid are not discriminated and the both are classified into fluid.
Crystal is a solid with the translational order.
Note that, the above definitions are not absolute,
since there are many substances which are difficult to be classified
such as glasses and liquid crystals, etc.
Instead, it is sometimes appropriate to define that the Alder transition itself
distinguishes fluid from solid.
In this section, we review briefly the definition of the translational order
in the crystal, and how it is inhibited  by the Mermin's theorem.

Consider a system with $N$ particles.
The density distribution function $\rho(\v{r})$ can be defined as
$$
\rho(\v{r}) = \sum_j^N \delta(\v{r} - \v{r}_j),
$$
with the positions of the particles $\v{r}_j$.
The fourier transform of the density $\hat{\rho}_{\v{k}}$ is given by
\begin{eqnarray}
\hat{\rho}_{\v{k}} &=& \int \mbox{d} \v{r} \exp{(i \v{k} \cdot \v{r})} \rho(\v{r})\\
&=& \sum_j^N \exp{(i \v{k} \cdot \v{r}_j)}. \label{eq_fourier}
\end{eqnarray}
If the system has the translational order, at least one reciprocal vector $\v{K}$.
With this wave vector, the value of Eq.~(\ref{eq_fourier}) becomes $O(N)$.
Therefore, the translational order parameter of this system is defined by 
\begin{equation}
\Psi = \frac{1}{N} \sum_j^N \exp(i \v{K} \cdot \v{r_j}).
\end{equation}
Mermin has proved that the translational order parameter $\Psi$ cannot exhibit
long range correlation in the two-dimensional system, provided that the 
the two-body potential $\Phi(r)$
satisfies the following conditions
\begin{equation}
\left\{
\begin{array}{cc}
\Phi(r) < & r^{-2+|\varepsilon|} \qquad (r \rightarrow \infty), \\
\Phi(r) > & r^{-2+|\varepsilon|} \qquad (r \rightarrow 0).
\end{array}
\right.
\label{eq_condition}
\end{equation}
The translational order $\Psi$ of two-dimensional system 
decreases faster than logarithmically as the system size increases, and will vanish
in the thermodynamic limit.
There are no such constraints in the three-dimensional system,
therefore, the nature of the Alder transition in two-dimensional system is
fundamentally different from that of the three-dimensional system.

The following points are worth noting. The Mermin's theorem cannot 
be applied to the self-gravity system, since the system does not
satisfy the condition (\ref{eq_condition}).
It is not trivial whether the theorem can be applied to the hard-particles,
since the hard-core potential is not differentiable.

\subsection{Bond-orientational order}
The constraint on the translational order in the two-dimensional system
is so week that some kinds of orderings are allowed to exist.
One of such the orderings is the bond-orientational order
which plays an important role in the KTHNY theory.
The six-fold bond-orientational order $\phi_6$ is defined as
\begin{equation}
\phi_6 = \ave{\exp(6 i \theta)}
\end{equation}
with the angle $\theta$ between a fixed axis and the bond connecting neighboring particles.
The average is taken for all pairs of neighboring particles.
The parameter $\phi_6$ becomes 1 when all particles are 
located on the points of the hexagonal grid, and it becomes $0$ 
when the particle location is completely disordered.
Therefore $\phi_6$ describes
how close the system is to the perfect hexagonal packing.
The neighbors in an off-lattice model are defined with the Voronoi construction.

Unlike the translational order parameter, the bond-orientational order parameter
does not change its value under uniform dilation of the system.
The long-range correlation of the bond-orientational order is not inhibited
in the two-dimensional system while the translational order is.

\subsection{Two kinds of defects}

The two-dimensional classical XY spin model cannot have
the spontaneous magnetization because of the Mermin-Wagner's theorem.
The system, however, involves the phase transition
involving divergence of the susceptibility~\cite{KT}.
This phase transition is called Kosterlitz-Thouless (KT) transition
which has some characteristic properties
such as the low temperature phase with power-low correlation and the essential singularity at the critical point.
The mechanism of this transition is explained by the unbinding of oppositely charged topological defects.

\begin{figure}[tbh]
\begin{center}
\includegraphics[width=0.3\linewidth]{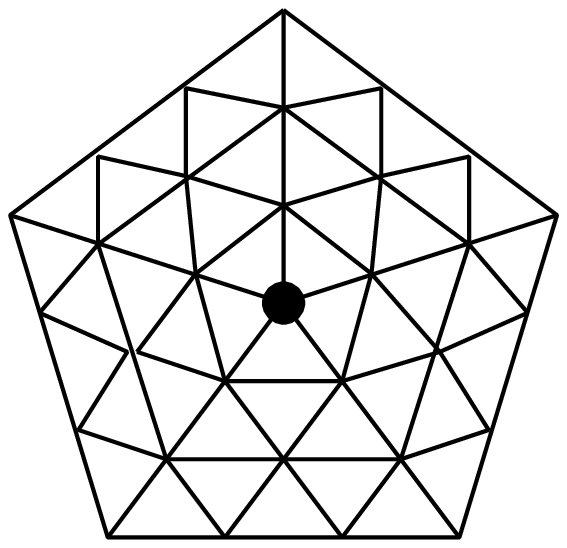}
\hspace{0.1\linewidth}
\includegraphics[width=0.3\linewidth]{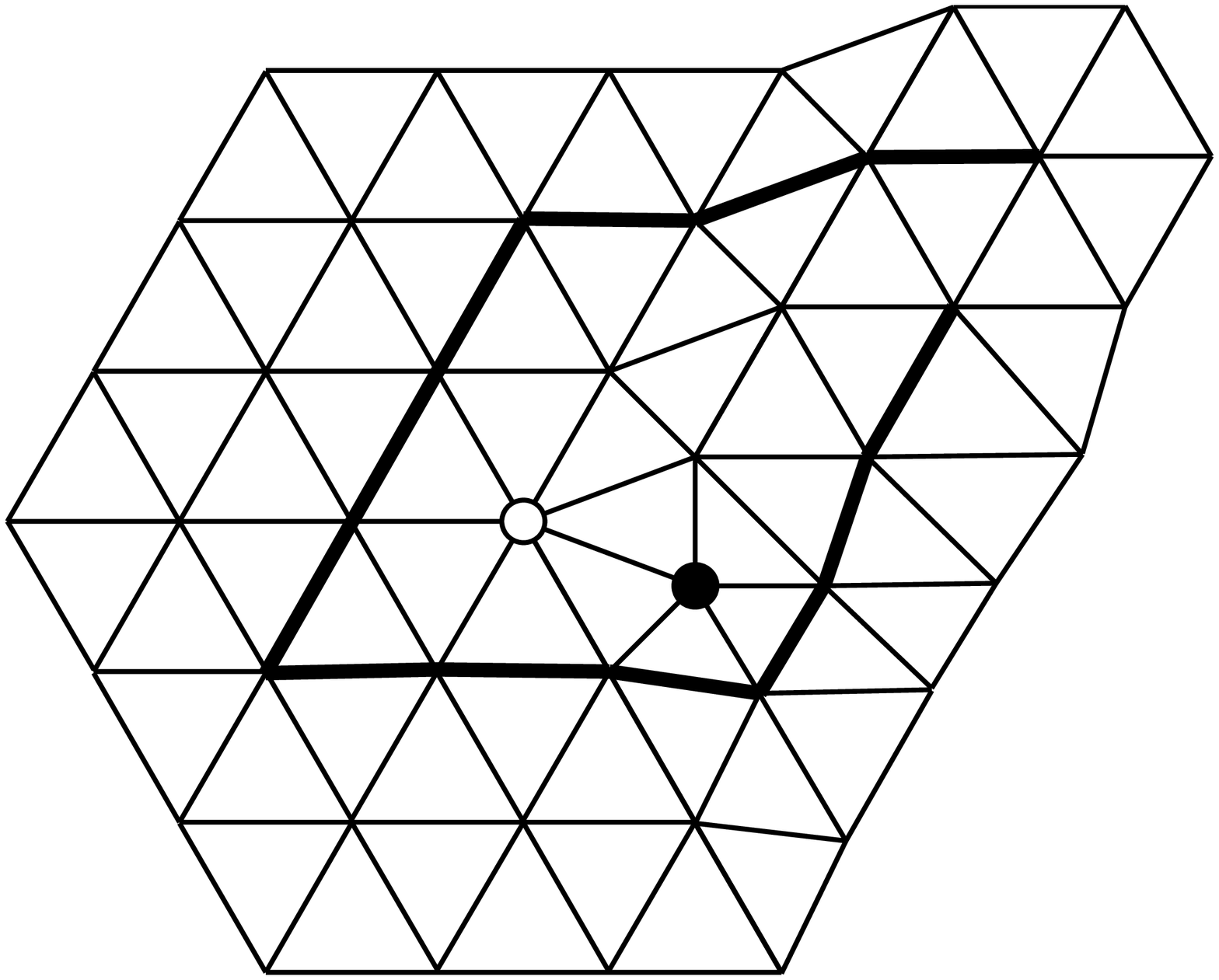}
\end{center}
\caption{
Defects in the triangular lattice.
(left) A disclination. The system has five-fold symmetry because of the disclination
while it has normally six-fold.
The charge of the disclination is defined by subtracting the number of the neighbors at the defect
from the average number of neighbors.
(right) A dislocation. The charge of dislocations is defined by
a Burgers vector which is a path around the defect fails to close.
Note that the dislocation can be described by the two disclination
with opposite charges.
In this figure, the dislocation consists of the 
positive (open circle) and negative (solid circle) disclinations.
}
\label{fig_defects}
\end{figure}

Similar unbinding mechanism can be applied to the two-dimensional melting.
In the two-dimensional crystal, there are two kinds of
defects which are dislocation and disclination.
Schematic drawings of the defects are shown in Fig.~\ref{fig_defects}.
While the dislocation destroys the translational ordering,
it does not destroy the bond-orientational ordering.
The disclination destroys the both ordering, and therefore, its influence is global.
Halperin, Nelson and Young have explained the two-dimensional
melting on the basis of the unbinding mechanism of the two kinds of defects
and this mechanism is now called the KTHNY theory.
The KTHNY theory predicts that the two kinds of orders involve
the KT transition independently.
As density decreases, solid melts into anisotropic fluid at density $\rho_{\mathrm{m}}$
and the anisotropic fluid becomes isotropic at the density $\rho_{\mathrm{i}}$.
Therefore, the solid becomes fluid via new phase which is called the hexatic phase.
The KTHNY theory is summarized in Table~\ref{tbl_kthny}.

\begin{table}[tbhp]
\begin{center}
\begin{tabular}{|c|ccc|}
\hline
Phase  & Solid & Hexatic & Fluid \\
\hline
Density & $  \rho > \rho_{\mathrm{m}} $ & $\rho_{\mathrm{m}} > \rho > \rho_{\mathrm{i}}$ & $\rho_{\mathrm{i}}  > \rho$ \\
Dislocation & Pair & free & Free \\
Disclination & Quartet & Pair & Free \\
translational order & quasi-long-range & short-range & short-range \\
Bond-orientational order & long-range order & quasi-long-range & short-range \\
\hline
\end{tabular}
\end{center}
\caption{
Summary of the KTHNY theory.
There is the hexatic phase between the solid and the fluid phases.
The quasi-long-range order means that 
the order decays to zero as a power of the distance.
The theory predicts that two order parameters involve the KT transition independently.
}
\label{tbl_kthny}
\end{table}

\subsection{First order vs. Continuous transition}

The KTHNY theory depends on some assumptions.
One of the important assumptions is that the
defects may be excited tenuously and uniformly.
If the defects concentrate locally, different arguments are required.
Chui calculated the free energy on the basis of the collective excitation of the defects,
and has concluded the first order transition~\cite{Chui}.
Whether defects are excited uniformly or collectively
is depends on the value of the core-energy of the effective Hamiltonian for defects~\cite{Saito1982}.
The core-energy plays a role of the chemical potential.
The KTHNY theory is justified for the large value of the core-energy,
and Chui's theory is justified in reverse~\cite{Strandburg1986}.
The value of core-energy depends on the details of the system,
and it is difficult to determine its value~\cite{Sengupta2000}.
Additionally, 
it is not trivial whether we can apply the elastic theory to the hard-disk system
while most of theories are based on the elastic theory.
Considering the simplified Bernal lattice,
Kawamura proposed theory without using the elastic theory
and has concluded the first order transition~\cite{Kawamura}.

\section{Non-equilibrium relaxation analysis}

\subsection{Finite-size effect on relaxation}

\begin{figure}[tbh]
\begin{center}
\includegraphics[width=0.48\linewidth]{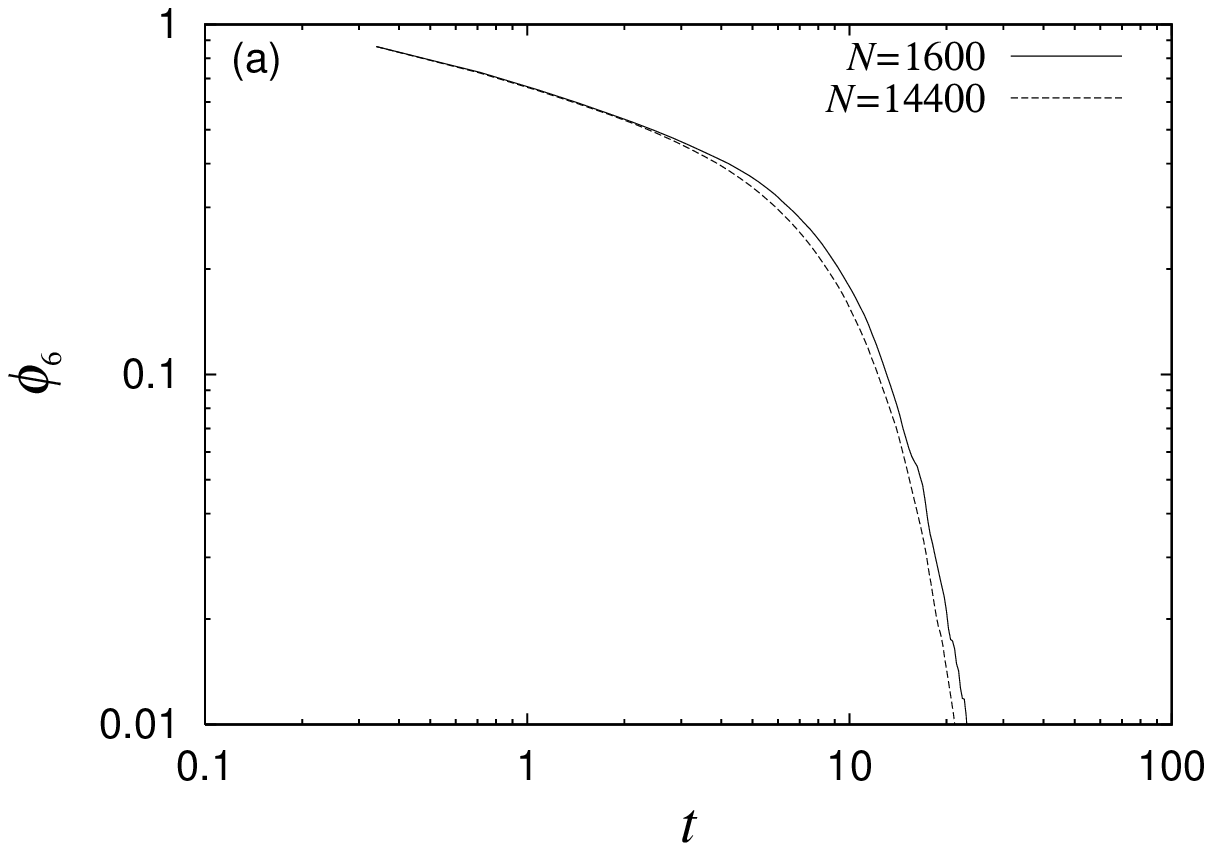}
\includegraphics[width=0.48\linewidth]{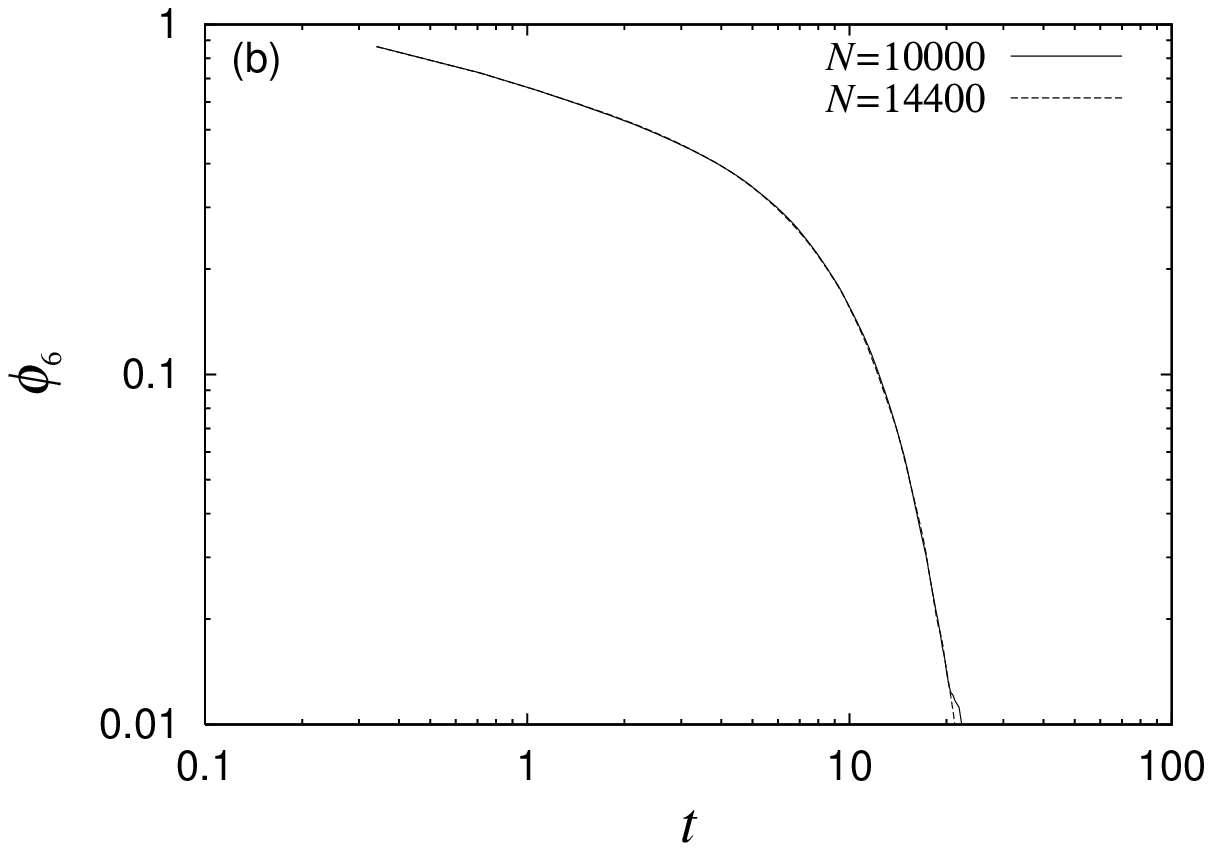}
\end{center}
\caption{
Finite-size effect on the relaxation.
Time evolutions of the bond-orientational order parameter $\phi_6$
are shown for the systems of $N=1600, 10000$ and $14400$.
(a) Comparison between the small and the large systems.
The behaviors begin to differ from around $t=2$.
(b) Comparison between larger systems.
The behaviors are completely equivalent, and therefore, this behavior
can be regarded as that in the thermodynamic limit.
}
\label{fig_fsize}
\end{figure}

Most of the numerical works studying the Alder transition
have used the equilibrium Monte Carlo (MC) simulations.
While the MC method is popular and has been widely used in statistical physics,
this method sometimes faces difficulty in achieving the equilibrium
state for the system with slow relaxation.
It is found that the time to achieve the equlibrium for the hard-disk system 
becomes much longer as the system size increases~\cite{Zollweg1992}.
Therefore, we introduce a new method, called the non-equilibrium relaxation (NER)
method, in order to avoid the finite-size effect~\cite{NERReview}.
The main idea of the NER method is to utilize the relaxation behavior to the equilibrium state.
At the beginning stage of the relaxation, the behavior of the system is
not influenced by the finite-size effect.
As time advances, the correlation length grows.
Therefore, the behavior of the system can be regarded as that of the 
thermodynamic limit until the correlation length reaches the system size.
In order to see this finite-size effect on the relaxation, 
time evolutions of the order parameter of the system with several sizes are shown in Fig.~\ref{fig_fsize}.
The time evolution of two different systems are equivalent
for a certain period of time, and this time becomes longer as the size of systems becomes larger.
Conversely, the relaxation behavior can be regarded as that of the system in the thermodynamic limit
provided the behavior are shared in two systems with different sizes.

\subsection{Dynamic scaling anaysis}

In order to investigate the phase transition from the relaxation behavior,
we study the dynamic scaling behavior of the order parameter.
We observe the time evolution of the bond-orientational order 
parameter of systems which initial configurations are set to be the
perfect hexagonal-packed.
The bond-orientational order parameter $\phi_6(\rho,t)$ 
is a function of density and time, and starts relaxing from $1$ to the value at the equilibrium.
Near the criticality, the behavior of $\phi_6(\rho,t)$ can be scaled as
\begin{equation}
\phi_6(\rho,t) \sim \tau^{-\lambda} \bar{\phi_6} (t/\tau), \label{eq_dynamic_scaling}
\end{equation}
with relaxation time $\tau$ and a density-independent parameter $\lambda$.
The scaling function is denoted by $\bar{\phi_6}$.
In the KT transition, the correlation length diverges
exponentially as 
\begin{equation}
\xi = b \exp{(a/\sqrt{\varepsilon})} \quad (\varepsilon \equiv (\rho_{\mathrm{i}} - \rho)/\rho_{\mathrm{i}}).
\end{equation}
According to the dynamics scaling hypothesis,
the relaxation time is associated with the correlation length as
\begin{equation}
\tau \sim \xi^z,
\end{equation}
with the dynamic scaling exponent $z$.
Finally, the divergence behavior of the relaxation time $\tau$ is expected to be
\begin{equation}
\tau(\varepsilon) = B \exp{(A/\sqrt{\varepsilon})}.
\label{ktscale}
\end{equation}
Therefore, we can determine the critical point $\rho_{\mathrm{i}}$
by observing the relaxation of $\phi_6$ at several densities~\cite{NERKT}.

The asymptotic behavior of $\phi_6(t)$ and its fluctuation at the critical point
are expected to be
\begin{eqnarray}
\phi_6(t) &\sim& t^{-\eta/2z},\\
N \left[
\frac{
\left< \phi_6(t)^2 \right>
}{
\left< \phi_6(t) \right>^2
} -1
\right]
&\sim& t^{d/z}.
\end{eqnarray}
with the critical exponents $\eta$ and $z$, and the dimensionality $d=2$~\cite{NERFluc}.
Therefore, we can determine the values of the critical exponents
by observing fluctuation behavior of the order parameter at the criticality.

\subsection{Numerical Results}

We perform the event-driven molecular dynamics (MD) simulation 
in order to observe the time evolutions of the bond-orientational order parameter.
The particle number $N$ is fixed at $23288$ throughout the following results.
The periodic boundary conditions are taken for both directions of 
the simulation box. Up to 512 independent samples are averaged at each density.

\begin{figure}[tbh]
\begin{center}
\includegraphics[width=0.48\linewidth]{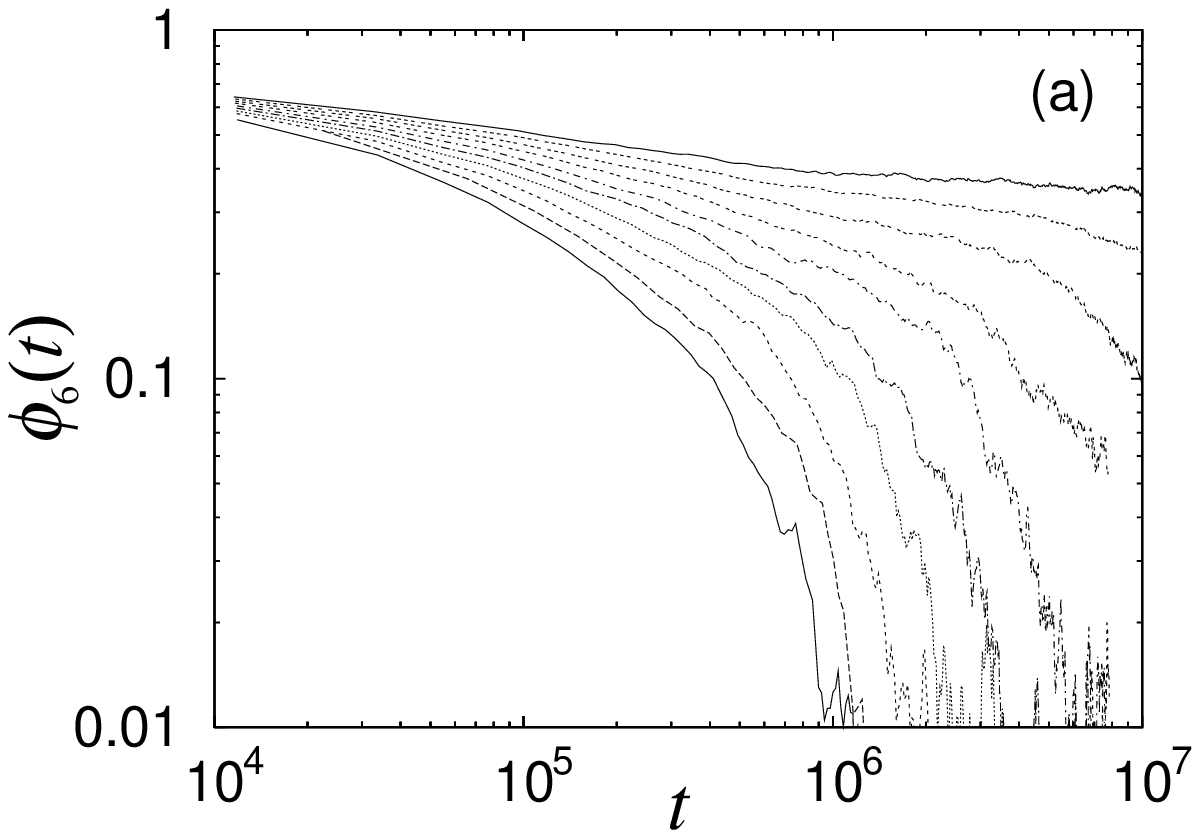}
\includegraphics[width=0.48\linewidth]{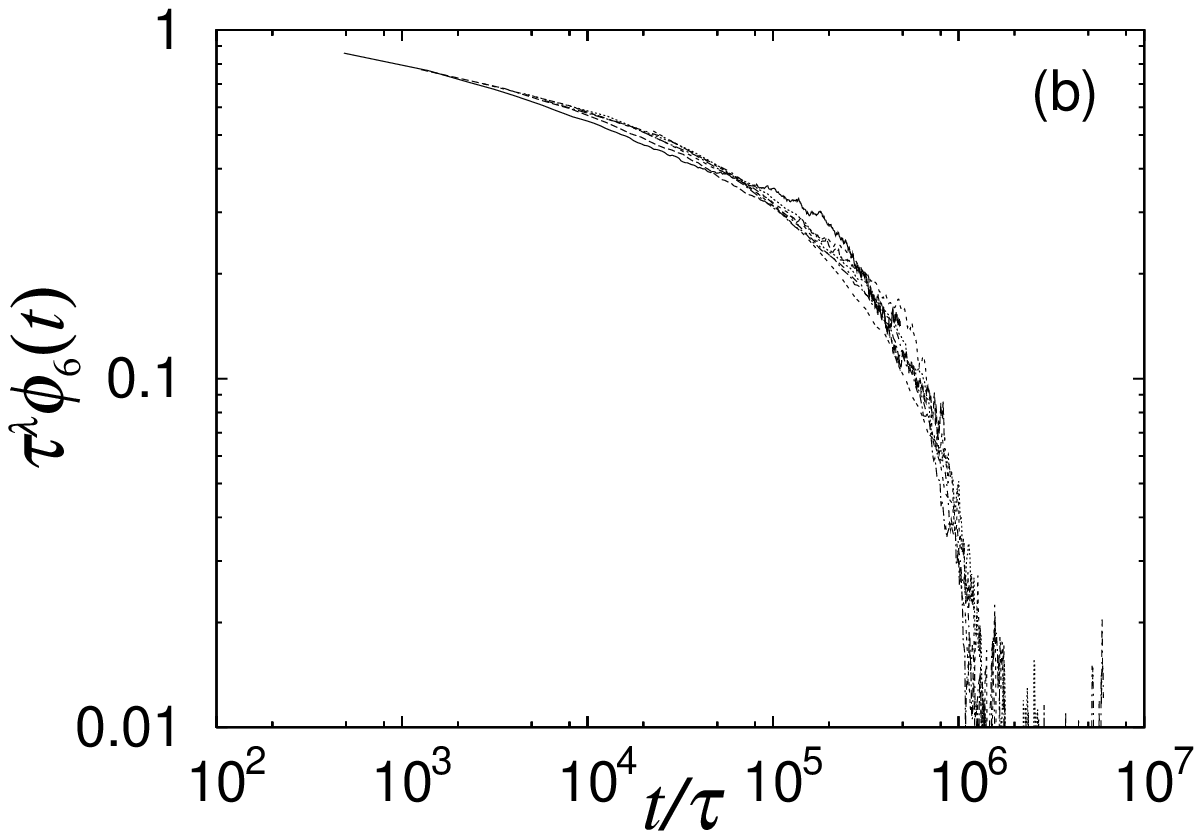}
\end{center}
\caption{
(a) Relaxation of the bond-orientational order $\phi_6(t)$
for various densities from $\rho = 0.878$ to $0.896$
Natural logarithms are used for the both axes.
(b) Scaling plot of bond-orientational order parameter
with appropriately chosen $\tau(\epsilon)$ and $\lambda$.
}
\label{fig_scaling}
\end{figure}

The time evolutions of the bond-orientational order parameter
are shown in Fig.~\ref{fig_scaling}(a) and the scaling plot
is shown in Fig~\ref{fig_scaling}(b).
From the divergence behavior of the relaxation time,
the critical point is determined to be $\rho_{\mathrm{i}} = 0.893(1)$.
The critical exponents at the critical point are determined 
to be $z=2.5(2)$ and $\eta = 0.25(2)$~\cite{Watanabe}.

\begin{figure}[tbh]
\begin{center}
\includegraphics[width=0.48\linewidth]{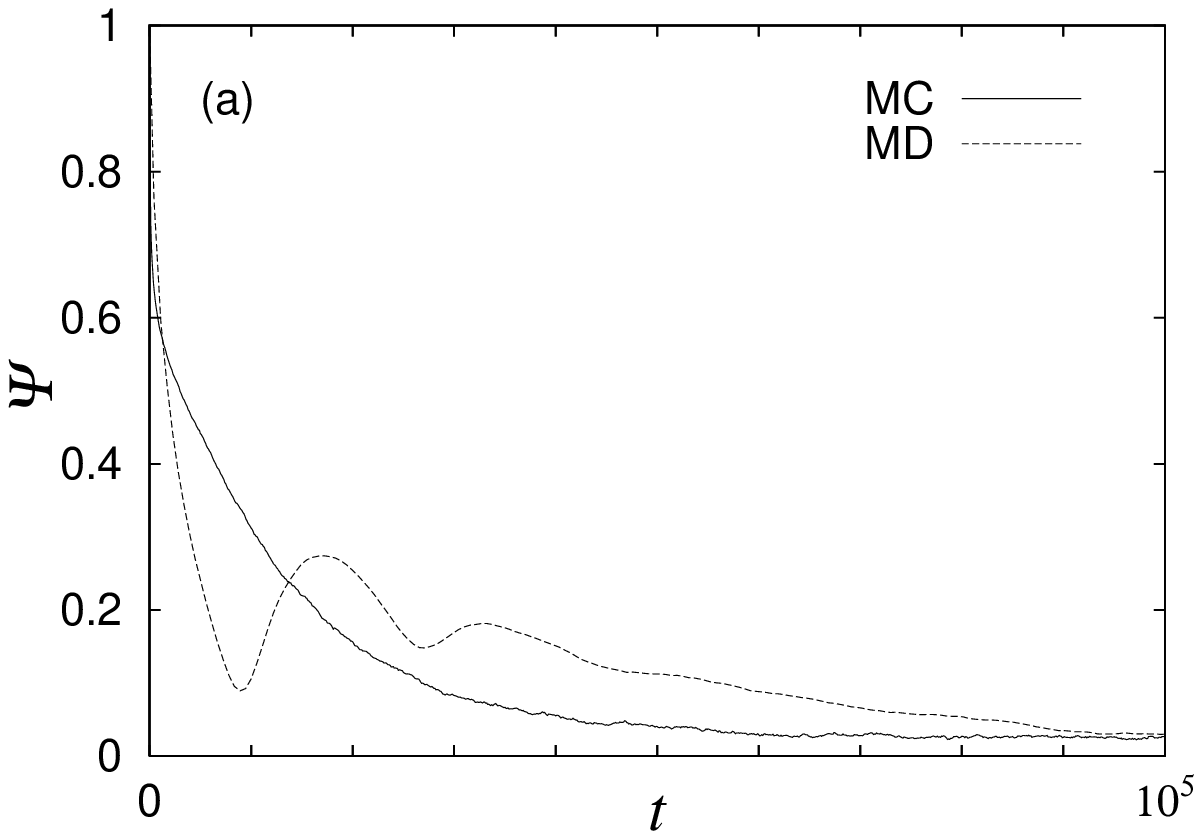}
\includegraphics[width=0.48\linewidth]{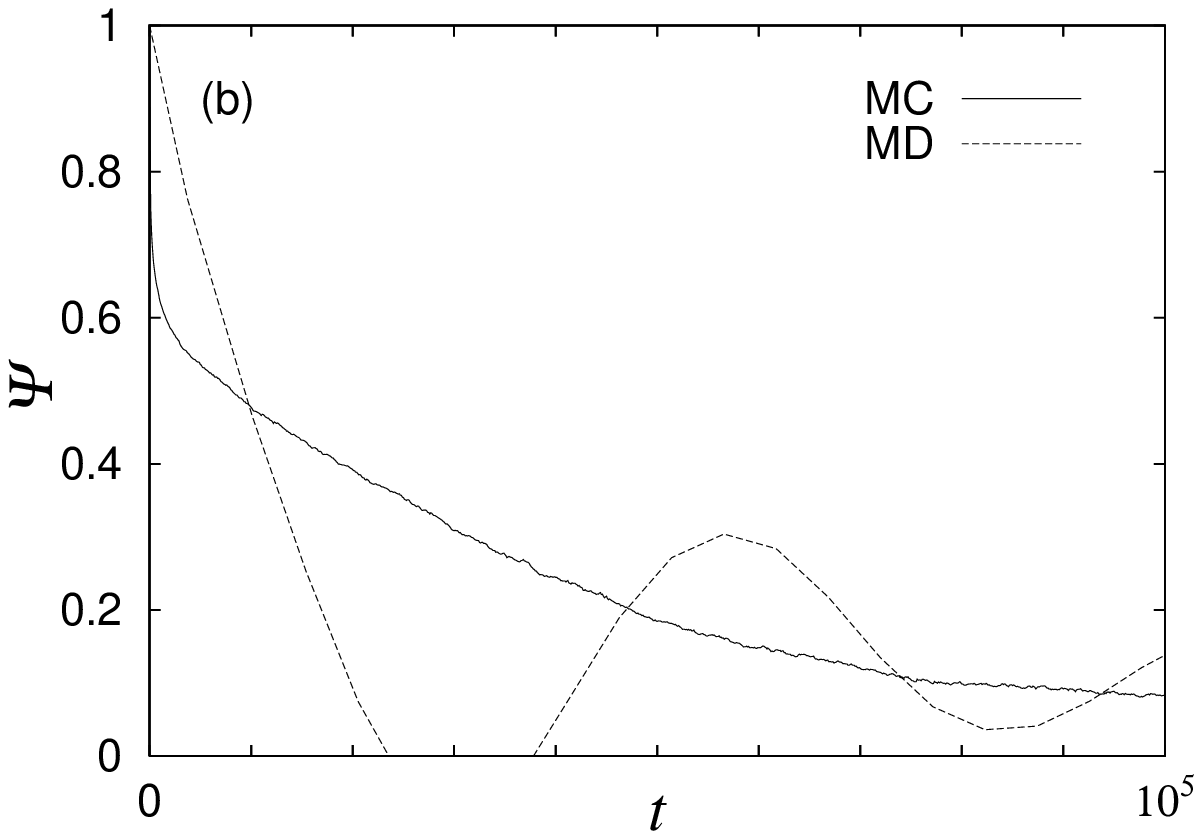}
\end{center}
\caption{
Comparison of the behavior of the translational order parameter
of the systems which time evolution are performed by
MC and MD.
The density is $\rho = 0.890$ in (a) and $0.900$ in (b).
Despite many independent samples are averages,
the results from MD have strong oscillation.
The amplitude of the oscillation becomes larger as density increases.
}
\label{fig_pos}
\end{figure}

\section{Summary and Further issues}

We have reviewed the Alder transition and the KTHNY theory.
The presented numerical results are consistent with the prediction of 
the KTHNY theory. Especially, the obtained value of $\eta \sim 1/4$ 
is a strong piece of evidence that the transition is of the KT type.
We have not studied the transition between the hexatic and the solid phase,
since the translational order parameter has strong oscillation
when the time evolution is performed by MD (see Fig.~\ref{fig_pos}).
This oscillation is caused by the conservation of the momenta,
therefore, the results from MC are free from it.
It is one of the further issues to investigate whether 
similar technique presented in this paper can be applied to MC method.

The studies on the Alder transition continues today
from the both sides of numerical works and experiments.
To the best of our knowledge, the largest simulation to date
contains $4 \times 10^{6}$ particles~\cite{Mak2006}.
While the scaling analysis of this study has supported the KTHNY theory,
the possibility of the week first order transition has also 
suggested from the pressure-density diagram.
The experiments of the two-dimensional melting have
been carried out on various systems, e.g., liquid crystals,
absorbed gas on graphite, two-dimensional plasma, and so on.
Recently, the excitation behavior of the disclinations 
have been directly observed in the dusty plasma~\cite{Quinn2001}.
In this experiments, the disclinations did not
get dissociated even in low enough density.
Instead, the grain boundary excitation was observed,
and therefore, the theory by Chui is supported.

Generally speaking, it is very difficult to 
distinguish the continuous transition from the week first-order transition.
Some new methods beyond the analysis of the scaling and the pressure-density diagram
have been waited such as the level spectroscopy method
which has achieved success to investigate the KT transition of the lattice systems~\cite{Nomura1995}.

\section*{Acknowledgements}
This work has involved collaboration with 
N. Ito, Y. Ozeki and S. Yukawa.
The author thank S. Miyashita and S. Todo for 
fruitful discussion.
Numerical works were carried out at the Supercomputer Center,
Institute for Solid State Physics, University of Tokyo and 
CP-PACS at the Center for Computational Physics, University of Tsukuba.
This paper is supported by the Ministry of Education, Science, Sports and Culture,
Grant-in-Aid for Young Scientists (B), 19740235, 2007, 
and for Scientific Research (C), 19540400, 2007.

%

\end{document}